# Anemometer-free indoor calibration of Doppler lidar using fiber optics and Monte Carlo simulation

**Andrew Black[1], Klaus Franke[2], Davide Trabucci[2], Pierre E. Allain[1], Lisa Lopez[1], Axel Albers[2]**

[1] Vaisala France SAS, 91400 Saclay
[2] DeutscheWindGuard GmbH, 26316 Varel

*Author to whom any correspondence should be addressed.

**E-mail:** andrew.hastingsblack@vaisala.com

**Keywords:** Doppler lidar, calibration, wind energy, Monte Carlo methods, turbulence box

## Abstract

Existing methods for calibrating and classifying Doppler lidars for wind energy applications (IEC 61400-50 series) are time-consuming and overestimate lidar measurement uncertainties. These shortcomings are due to the reference instruments: anemometers on meteorological masts. The input signals in field calibrations are real wind conditions, and include large uncertainties associated with the meteorological mast and surrounding terrain.

This research presents validation of a new methodology for lidar calibration. The new approach has two steps. First, the uncertainty of the lidar line of sight velocity measurements (LOS) are measured. Second, those intermediate uncertainties are used to derive the uncertainty of the horizontal wind speed using a simulated wind field.

Intermediate uncertainties are assessed using a fiber optic bench as a synthetic wind tunnel called Simulation of the Atmosphere with Fiber Optics Multi-Velocity (SAFO-MV). Using SAFO-MV, the lidar's LOS is shifted using an acousto-optic modulator (AOM) and backscattered from a long fiber spool, mimicking backscatter from a uniform, low turbulence wind field. The reference signals are SI-traceable via calibration of the AOM's RF-driver. Sources of uncertainty in the bench and the laboratory process are examined and found to be small.

Horizontal wind speed uncertainties are derived via Monte Carlo uncertainty propagation. Wind fields are simulated using PyConTurb, an implementation of the KSEC turbulence model. A virtual lidar in the simulation is shown to replicate lidar sensitivities derived by IEC 61400-50-2 classification campaigns. The replication of known device sensitivities and the traceability of the intermediate signals comprise a validation of the measurement model, enabling its use for industrial lidar calibration

The proposed routine yields uncertainties comparable to current-art IEC lidar calibrations, while significantly reducing the calibration time. This new technology has the potential to streamline lidar deployments and reduce uncertainty in wind resource assessment.

## 1. Introduction

Doppler wind lidar data is used for two primary industrial applications in wind energy: energy yield assessment (EYA), measurement campaigns used for wind farm development; and, power performance testing (PPT), contractual validation tests of wind turbine performance. The lidar measurements used in these two applications are traceable to international standard measures of wind speed. These standards are defined by expert groups in the International Electrotechnical Commission (IEC), specifically the technology standard IEC 61400-50-2 (vertically profiling lidar), 50-3 (nacelle-mounted lidar), and 50-4 (floating lidar, FLS), as well as use case standards such as IEC 61400-12-1 (PPT) (1) (2) (3) (4).

All these use cases rely on calibrated cup anemometers as reference instruments to derive SI-traceable uncertainties. As such, when lidar are used in these wind energy applications, they inherit the uncertainties of the cup anemometers. This inheritance has been widely criticized as the source of systematic overestimation of lidar measurement uncertainty at wind energy industry conferences (5), in industry working groups (6), and in joint-industry projects (7). In response to the absence of an industrialized, traceable technique for lidar uncertainty estimation, a variety of new technologies have been developed. In (8) a method using a small continuous-wave (CW) lidar or "lidic" is demonstrated for calibrating nacelle-mounted lidar. Another method using a bi-static lidar is demonstrated in (9). In each of these techniques, the reference instrument is a novel lidar with a form factor allowing for deployment into a calibrated wind tunnel, thus eliminating the need for a cup anemometer in the traceability chain.

Cup anemometers are equivalently used for PPT and EYA in the wind energy industry and share many methodologies for uncertainty estimation. Calibration of cup anemometers is carried following standards developed by the IEC (10) using wind tunnels that are certified and operated following ISO 3966:2025 (11). In *Dahlberg* (12), a method for anemometer sensitivity analysis was demonstrated, using laboratory measurements of an anemometer's response as an input to a numerical model. Using this Monte Carlo uncertainty model, *Dahlberg* derived sensitivities and uncertainties across a wider range of environmental variables than possible in field or wind tunnel calibration. This methodology is incorporated to the IEC Standard (10), demonstrating the usefulness and industrialization of model-based or Monte Carlo methodologies for wind speed calibration and sensitivity analysis. Monte Carlo methods follow principles of metrology described in (13), while many methods in wind measurement IEC and ISO standards follow JCGM's *Guide to the expression of uncertainty in measurement* (14).

Lidar simulations, or "virtual lidar", are used today in remote sensing and wind energy research for development of feed-forward wind turbine control systems (15), for planning lidar measurement campaigns (16), and for development of new algorithms and lidar measurement strategies (17) (18). In virtual lidar, elements of lidar physics and the atmospheric physics are simulated to replicate the expected wind measurements. The virtual atmosphere is generated either via a turbulence box (15) (19) or a large-eddy simulation (18) (20). The virtual lidar measurements may be represented as point measurements at the appropriate location and angle, as weighted averages along the lidar beam line of sight or using simulation of individual pulses interaction with atmospheric turbulence (21).

## 2. Method

In this research, a new SI-traceable calibration technique for wind speed measurements using pulsed, Doppler wind lidar is presented. This new method follows the paradigm of Monte Carlo uncertainty estimation, using laboratory measurements of the lidar's response as an input to a numerical model. The derivation of the lidar response is carried out using a novel, fiber optic bench called SAFO-MV (Simulation of the Atmosphere with Fiber Optics Multi-Velocity) (22). The bench components used to control the wind speed signal are calibrated by ISO17025-certified laboratories, the first link in the SI-traceability of the new approach.

The laboratory-derived responses of the lidar are used as an input to a numerical model to derive the uncertainty and environmental sensitivities of the lidar. The numerical model is based on an open-source code, MoCaLUM (Monte Carlo Lidar Uncertainty Model) described in (23). This code enables repeated simulations of realistic flow fields using PyConTurb, a Python implementation of the KSEC model of turbulence (24). A complete list of model inputs is included in Section 2.3. The numerical model may output either uncertainty estimates of the horizontal wind speed derived from the distributions of error in comparison to a virtual point reference, or sensitivities to a broad range of environmental conditions, which are part of the model configuration. Both use cases are presented.

### 2.1. Fiber Optic Laboratory Calibration of Lidar-Measured Velocities

The SAFO-MV bench simulates the key mechanisms that Doppler lidar use to measure wind velocities in the atmospheric boundary layer (ABL). A glass fiber spool generates diffuse backscatter via Rayleigh scattering. This backscatter effectively emulates the backscatter from motionless, randomly distributed aerosols in the atmosphere. The backscatter is primarily caused by glass defects, and silica atoms in the glass fiber. In the atmosphere, the aerosols are not motionless, but advect with the wind, generating a Doppler shift in the backscattered laser light. This difference requires another component: the Doppler shift is implemented in the bench by shifting the outbound and inbound lidar signals using an acousto-optic modulator. These are the two key characteristics of pulsed Doppler lidar measurement principle: spatially diffuse, Doppler-shifted backscatter. In the same way that a wind tunnel emulates drag in a controlled environment to activate the measurement mechanism of cup anemometers, the SAFO-MV bench emulates the key elements of pulsed Doppler lidar measurement mechanism.

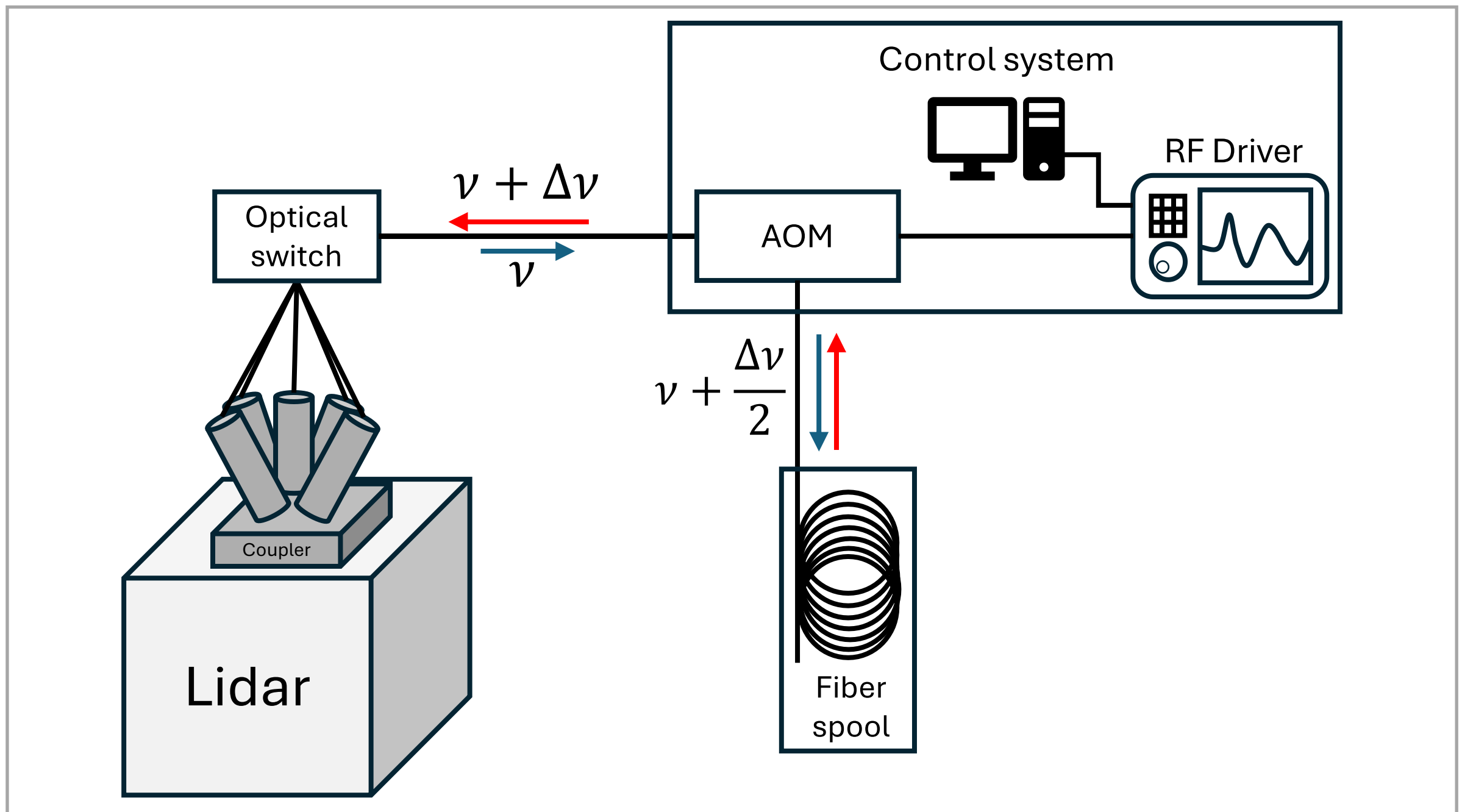

**Figure 1:** Diagram of SAFO-MV fiber optic laboratory bench. The bench consists of (1) an optical coupled with five custom telescopes positioned at the lidar window, connected to (2) an optical switch which selectively redirects the light to a central fiber connected to the bench SAFO-MV. Within the bench there is (3) a modulator, controlled by a radio frequency (RF) driver, to modify the power and the frequency of the laser light, simulating both carrier-to-noise ratio (CNR) and wind velocity, and (4) a fiber spool, 1000 m length, which generates backscatter. The system is controlled by a computer running software for the SAFO-MV bench, allowing the user to program sweeps over lidar line-of-sight (LOS) beams, CNR, and radial wind speed

The bench is fibered up to an optical head shown in Figure 2, which is positioned at the lidar window, including five telescopes that focus the light on five fibers connected to an optical switch. The length of the optical fiber sets the maximum range under study and thus can be easily customized. The lidar system operates under its normal configuration (Table 1) while the SAFO-MV control system sweeps over frequencies corresponding to the range of velocities and carrier-to-noise ratios (CNR) corresponding to those observed in the ABL. The resulting measurements, recorded by the lidar itself, are compared against the programmed Doppler shifts recorded in the SAFO-MV control software. The system accumulates lidar measurements for 650 seconds in each configuration. The software filters any erroneous data during transitions between bins and computes the average and standard deviation for each CNR, LOS and velocity bin.

The RF drivers used to modulate velocity and CNR are calibrated by ASERTI Metrology, a COFRAC-accredited laboratory with expertise in time-frequency domain calibration. This calibration serves as the metrological traceable link to the International System of Units. The uncertainty of the RF driver calibration is negligible (5.95622 x $10^{-8}$ %). The frequency modulation, in megahertz, applied to the lidar pulse as it enters and exits the fiber spool, is directly proportional to velocity shift measured by the lidar in meters per second via the Doppler effect.

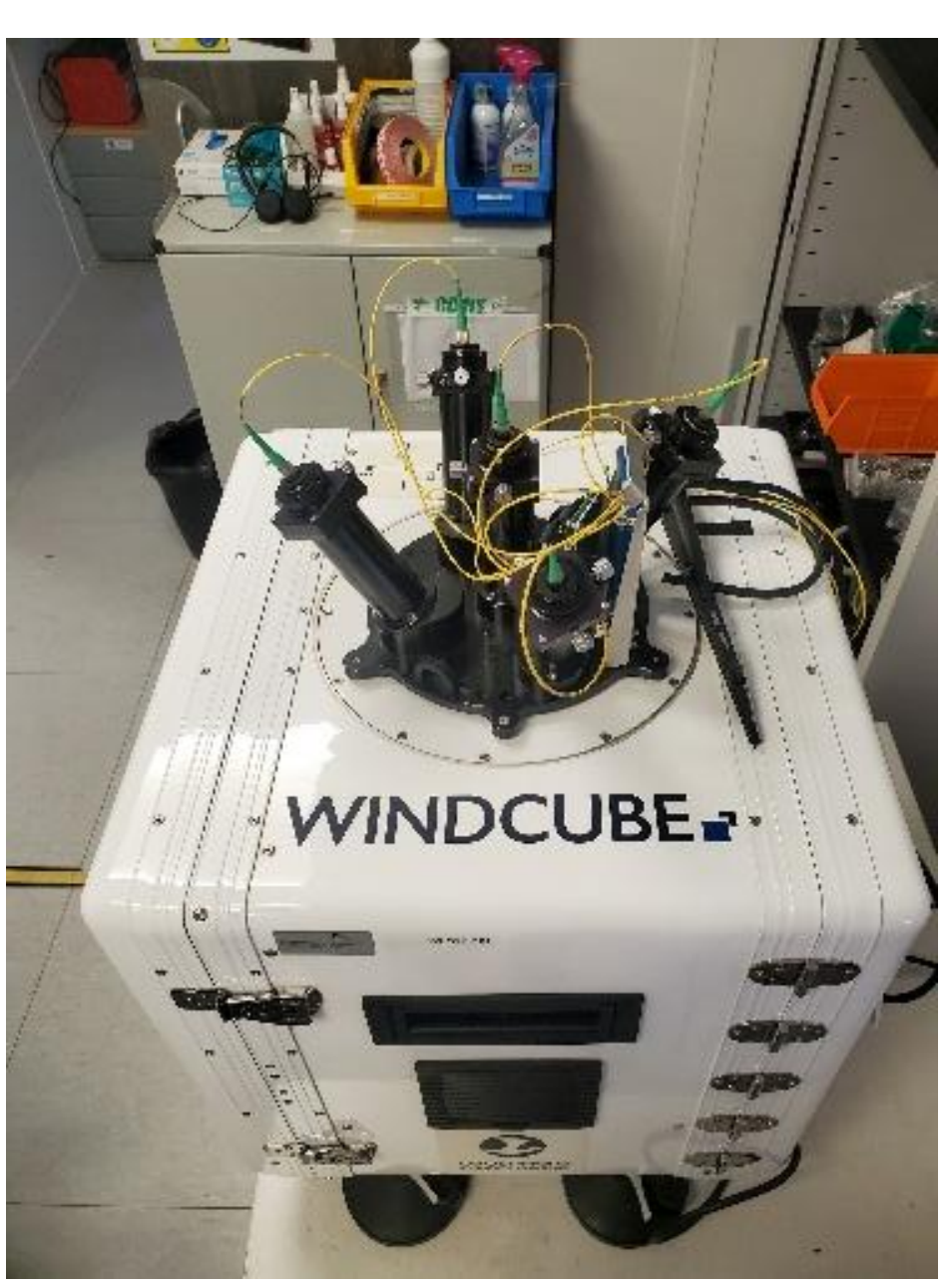

**Figure 2:** Optical coupler from SAFO-MV mounted on a WindCube lidar

SAFO-MV estimates of radial velocity uncertainty rely on the assumption that the lidar's CNR completely models the aerosol density in the atmosphere. If the CNR of measurements in the SAFO-MV bench is similar to those observed in the atmosphere, then the glass impurity backscatter and aerosol backscatter may be treated equivalently. CNR is the ratio between the backscattered signal power and the noise power within a single range gate. It is measured using the Doppler spectrum after the spectral processing, but before incoherent accumulation of spectra. It quantifies the instantaneous detectability of the Doppler signature and is independent of the source of the backscatter provided that the optical path and detection chain are identical. In the case of SAFO-MV, the optical path and detection chain are identical, as the lidar itself, operating with standard configuration, is a part of the bench apparatus. However, the assumption of equivalence is limited by several factors. The fiber backscatter is uniform throughout the spool. In the atmosphere, aerosol density can change significantly in the presence of clouds or fog, or in large turbulent structures such convective plumes or Kelvin-Helmholtz waves (25). Another difference is variation of CNR within the typical 10-minute sampling period. During a frontal passage or other mesoscale weather phenomena, the CNR can change abruptly within the sample period due to changes aerosol density in the different air masses. In SAFO- MV and the subsequent simulations, the CNR is modeled as uniform within the 10-minute period.

**Table 1.** Lidar and SAFO-MV configuration during calibration run. The lidar configuration used is identical to that used by default in IEC-compliant wind energy applications and in the device type Classification. Beam angles are set by fixed telescopes. Fields denoted (*) are associated with the SAFO-MV control system. Velocity is modulated in 1 m/s steps. CNR modulated in 2 dB steps.

| Configuration | Value |
|---|---|
| Pulse length | 25 m |
| Pulse shape | Gaussian |
| Sampling rate | 1.25 Hz |
| Laser frequency | 1.55 µm |
| Beam angles (polar, azimuthal) | (62°, 0°), (62°, 90°), (62°, 180°), (62°, 270°), (0°,0°) |
| Ranges (m) | 80, 90, 100, 110, 120, 130, 140, 150, 160, 170, 180, 190, 200, 250, 300 |
| Scan mode | Doppler beam swinging |
| Velocity range* | -18 m/s to 18 m/s |
| CNR range* | -28 dB to -6 dB |
| Accumulation time* | 650 seconds |

CNR is modeled as uniform within the 10-minute period. Finally, the atmospheric CNR profile is not uniform along the line of sight: it varies with range due to the lidar focus and the aerosol distribution. These limitations represent edge cases, outside the envelope of validity of a SAFO-MV calibration. Nonetheless, SAFO-MV and the accompanying simulation capture the key drivers of device uncertainty and bias in ordinary atmospheric conditions.

### 2.2. Fiber Optic Laboratory Calibration of Distance and Range-Weighting Function

Every pulsed lidar requires careful calibration of the outbound pulse's zero distance ($Z_d$). This calibration is critical for lidar ranging: the pulse timing determines the altitude of the velocity measurement. Similarly, the pulse shape determines the relative weighting of radial wind speeds within the measurement volume. Even with a perfect calibration of $Z_d$, asymmetrical pulses can generate small velocity biases.

In (22) a fiber optic bench and process are described which perform these two calibrations for Doppler lidar. This process is used by factory technicians to determine the exact time of flight to the window (TOFTW) and to record the shape of the reflected pulse. To derive the range-weighting function (RWF) the lidar's apodization function is convolved with the pulse shape, in intensity, as shown in Equation 1. The apodization function is derived from a Gaussian window and is used to

$$RWF(z) = f_{pulse}(z) \otimes f_{apod}^{2}(z) \quad (1)$$

preprocess the intensity time series before applying a Fast Fourier Transform to the backscattered signal. The RWF describes how the lidar weights contributions along the line of sight within a given range gate. It results from the convolution of the pulse envelope with the temporal gating window, squared, and sets the effective probing volume as well as the spatial averaging of the measured radial velocity (21). Both $Z_d$ and the RWF are recorded for each system and included in the device metadata for measurement traceability.

As mentioned in the previous section, CNR is not uniform along the line of sight. The range-dependence of the CNR profile is folded into a CNR-weighted RWF, which sets how velocity contributions are averaged within the gate:

$$RWF(z) = f_{pulse}(z) \otimes f_{apod}^{2}(z) \times 10^{(CNR(z)/20)} \quad (2)$$

RWF in the presence of CNR gradients are shown in Figure 3, alongside typical gradients observed at the Saclay, France manufacturing facility over the course of one year. This modelisation of the RWF is first presented here, and described in further detail in (21)

### 2.3. Monte Carlo Simulation

The Monte Carlo simulation is based on the "Monte Carlo Lidar Uncertainty Model" or MoCaLUM, an open source lidar uncertainty estimator developed in (23). The code leverages PyConTurb ("Python Constrained Turbulence") an open source turbulence box generator used to simulate realistic flow fields for wind turbine load testing and other wind energy applications (24). MoCaLUM embeds a virtual lidar within a PyConTurb turbulence box. The Monte Carlo uncertainty model is implemented via repeated runs over a broad range of atmospheric conditions and lidar configurations. In this research, the virtual lidar is configured using the output of the SAFO-MV bench, including (1) mean and (2) standard deviation of radial velocity as a function of radial velocity and CNR, and (3) $Z_d$ and (4) RWF.

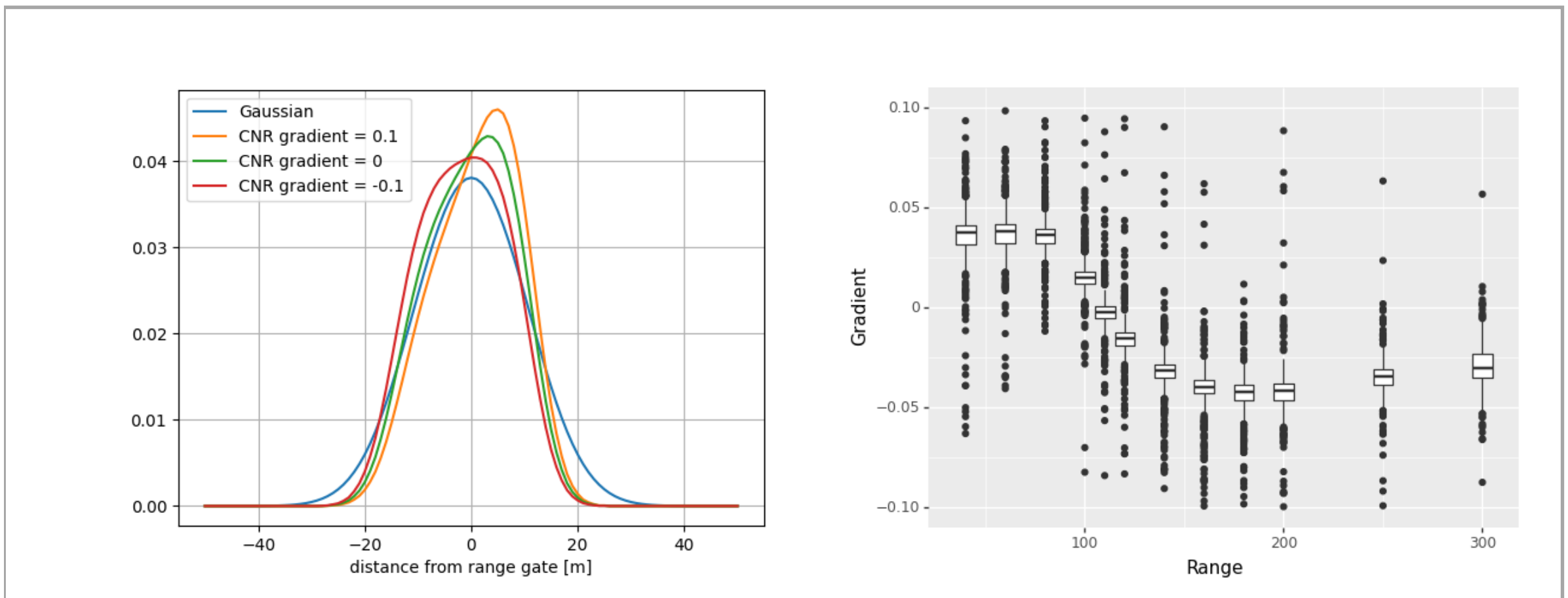

**Figure 3:** (Left) RWF for three CNR gradients following Equation 2. ±0.1 dB/m are extrema. Blue curve is a pure Gaussian, added for reference. Note the asymmetry of the RWF even with a CNR gradient of zero, originating in a slight pulse asymmetry. (Right) CNR gradients from WindCube v2.1 WLS7-1334 in Saclay, France during 2023. Note the gradient's zero crossing at 110 m, aligned with the lidar focal distance.

**Table 2:** Input distributions and configurations to individual 10-minute simulations

| Term | Distribution | Description | Origin |
|---|---|---|---|
| $\sigma_u(y,\ z,\ I_{ref}, V_{hub})$ | Gaussian | Longitudinal component of turbulence (m/s). Uniform over y and z. $I_{ref}$ set by user (time series from field campaign). $I_{ref}$ is reference turbulence intensity (%) and $V_{hub}$ is the hub height wind speed (m/s). Turbulence follows a Kaimal spectrum with $\sigma_u$ as the integral. Exponential coherence with neighboring pixels via Cholesky decomposition. | PyConTurb |
| $\sigma_v(y,\ z, \sigma_u)$ | Gaussian | Transverse component of turbulence (m/s). Uniform over y and z. IEC 61400-1 Ed. 4. 6.3.1 Kaimal spectrum derived with $\sigma_v = 0.8 * \sigma_u$ as magnitude. Exponential coherence with neighboring pixels via Cholesky decomposition. | PyConTurb |
| $\sigma_w(y,\ z, \sigma_u)$ | Gaussian | Vertical component of turbulence (m/s). Uniform over y and z. IEC 61400-1 Ed. 4. 6.3.1 Kaimal spectrum derived with $\sigma_w = 0.5 * \sigma_u$ as magnitude. Exponential coherence with neighboring pixels via Cholesky decomposition. | PyConTurb |
| $I_{ref}$ | User-defined time series | Defined as $\sigma_{HWS,10\ mins}/\overline{Wind\ speed}_{10\ mins}$ | |
| $\sigma_{LOS}(CNR, RWS)$ | Gaussian | LOS uncertainty for lidar beam. Uniform in z. Derived from SAFO-MV bench. | SAFO-MV |
| $\mu_{LOS}(CNR, RWS)$ | Scalar | Bias of LOS wind speed as a function of CNR and RWS. Empirically derived from SAFO-MV bench. | SAFO-MV |
| Availability | Uniform | Removal of samples from the simulated 10-minute period simulating data gaps due to low CNR. Set to 100% in example. | MoCaLUM |
| CNR | User-defined time series | CNR used to select LOS uncertainty distribution. | MoCaLUM |
| CNR gradient | User-defined time series | CNR gradient within range gate at measurement height. | MoCaLUM |
| Wind speed | User-defined time series | Horizontal wind speed of turbulence box | MoCaLUM |
| Wind direction | Classification time series | Azimuthal orientation of turbulence box to lidar machine coordinate system | MoCaLUM |
| Vertical wind speed | Fixed | Vertical orientation of turbulence box to lidar machine coordinate system | MoCaLUM |

For each input distribution, the random number seed is recorded and incremented between successive trials in the Monte Carlo run. The x-, y-, and z-resolutions of the turbulence box are configured such that the four radial velocity measurements are not artificially correlated. The dimensions in each orientation must be below the minimum distances between the different beams. This minimum cell size is computed using the lidar beam tilt and azimuth angles, the measurement height, z, and the LOS sampling rate and period.

Using the minimum measurement height (40 m), the tilt and azimuth angles (62° and 90°), and the low extremum of wind turbine hub heights (80 m), a = 85.1 m, and b = 60.2 m. These represent the minimum cell size for 80 m altitude to ensure that none of the four beam locations use the same randomly generated value, the cross correlation between beams. To prevent artificial autocorrelation in consecutive samples of the same radial wind speed, the longitudinal cell size must ensure that consecutive samples from the same line of sight are unique, independent draws from the input distributions. The sampling rate of the lidar is included in this calculation. For the WindCube v2.1, the sampling rate for each individual beam is 0.25 Hz, for all beams together 1.25 Hz, corresponding to a period of 4 s for the five beams. In the MoCaLUM simulation, the vertical beam is ignored, as it is not relevant for horizontal wind speed uncertainty derivation. The sampling rate is therefore 1 Hz. In this simulation paradigm, the individual lidar pulses (~20 kHz) are not included, so this change does not affect the results. For a wind speed of 4 m/s, the minimum cut-in wind speed for wind turbines (and typical minimum speed calibrated for wind sensor) the distance covered is 4 m/s x 4 s = 16 m. This represents the maximum cell size to ensure that each consecutive sample at each line of sight corresponds to a unique random draw from the simulation distributions. Simulations shown in this research use an x- and y-grid cell size of 10 m.

Wind field reconstruction (WFR) is a class of algorithms that transform radial wind speeds into Cartesian ($u$, $v$, $w$) and cylindrical (*wind speed*, *wind direction*, *vertical wind speed*) reference frames. For the WindCube v2.1 and XP products, two different wind field reconstruction algorithms are applied to the LOS data, and they are averaged together, known as hybrid wind field reconstruction (26). The two reconstruction algorithms are scalar averaging and vector averaging. For four consecutive LOS, the scalar averaging WFR algorithm is as follows:

$$u_i = \frac{S_1 - S_3}{2\sin\varphi}$$

$$v_i = \frac{S_2 - S_4}{2\sin\varphi}$$

$$V_{scalar} = \frac{1}{600}\sum_{i=1}^{600}\sqrt{u_i^2 + v_i^2} \tag{3}$$

Where $S$ corresponds to a radial wind speed, and $\varphi$ is the lidar's opening angle. Equation 4 shows vector wind field reconstruction:

$$V_{vector} = \sqrt{\frac{1}{600}\sum_{i=1}^{600}u_i^2 + \frac{1}{600}\sum_{i=1}^{600}v_i^2} \tag{4}$$

The two algorithms can be understood as (scalar) reconstruct-then-average and (vector) average-then-reconstruct. Finally, these two are combined to perform hybrid WFR:

$$V_{hybrid} = \frac{1}{2}V_{scalar} + \frac{1}{2}V_{vector} \tag{5}$$

Hybrid WFR reduces the lidar's sensitivity to turbulence, and increases agreement with cup anemometry, the traditional reference instrument for wind energy applications (26). The hybrid weights can be equivalently expressed as $p$ and (1-$p$) with $p$ = 1/2. $p$ is a configuration parameter in the MoCaLUM program and the lidar firmware. Note that the WindCube v2.1 uses $p$ = 2/3, and the WindCube XP uses $p$ = 1/2

When gathering LOS samples in the turbulence box, a random value drawn from a Gaussian distribution is added to the measured radial velocity. The distribution is defined as $\mathcal{N}_{LOS,n}(\mu,\sigma)$ where $\sigma_{LOS,n}(CNR, RWS)$ is the random uncertainty derived from SAFO-MV, and $\mu_{LOS,n}(CNR, RWS)$ the systematic bias derived from SAFO-MV, with the mean and standard deviation of each distribution derived from the SAFO-MV bench calibration for each LOS individually.

### 2.4. Propagation of Uncertainties

The MoCaLUM simulation including the SAFO-MV, $Z_d$, and RWF calibrations capture many, but not all of the uncertainties of a pulsed, profiling lidar. Five additional uncertainties are included in the uncertainty budget, which cannot be appropriately modeled in the MoCaLUM software:

- Repeatability of the mean bias estimates from SAFO-MV
- Repeatability of the standard deviation estimates from SAFO-MV
- Standard uncertainty of SAFO-MV bench
- Uncertainty of lidar beam angles
- Uncertainty of $Z_d$ calibration

Reproducibility of mean and standard deviation are not possible to estimate yet, as this requires at least two independent laboratories to repeat the calibrations on identical systems.

To estimate the repeatability of mean and standard deviation, multiple operators carried out SAFO-MV calibrations of the same lidar system. Each operator followed the same process instructions, software, and bench in the same laboratory in the Vaisala facility in Saclay, France. The room is a temperature-controlled photonics R&D laboratory.

**Table 3:** Timeline of repeatability tests on WLS7-7008 WindCube lidar

| Operator | Test | Start date | End date |
|---|---|---|---|
| A | 1 | 2/21/2025 | 2/23/2025 |
| B | 2 | 2/24/2025 | 2/26/2025 |
| C | 3 | 2/27/2025 | 3/2/2025 |
| D | 4 | 3/4/2025 | 3/6/2025 |
| E | 5 | 3/7/2025 | 3/9/2025 |

Run 3 by Operator C was identified as an outlier due to an error in the mounting of the optical head and is excluded from the analysis. From the calibration data, we may estimate the reproducibility error, $s_W^2$, following ISO 5725-4 (2020) for within-laboratory variance (27). Estimates of $s_L^2$, the between-laboratory variance, are beyond the scope of this research, but may be estimated once additional laboratories use SAFO-MV benches for lidar calibration. Similarly, $s_r^2$, the mean of $s_W^2$ values derived at different laboratories is not included.

Relevant uncertainties are estimated via the mean deviation $\delta_{\ell\text{cv},i}$ and standard deviation $s_{\ell\text{cv},i}$ in each (LOS, CNR, velocity) bin (hereafter "$\ell$cv-bin") about the known applied frequency shift, with sufficient samples present at the end of the SAFO calibration period. The subscript $i$ indicates the operator ($i \in [1, 2, 4, 5]$).

To estimate $s_W^2$ from our experiment, we derive the grand mean of each value $\overline{\overline{\delta_{\ell\text{cv}}}}$ and $\overline{\overline{s_{\ell\text{cv}}}}$ in each $\ell$cv-bin. Next, we compute the standard deviation of the four reproducibility tests in each $\ell$cv-bin, $s_{\delta_{\ell\text{cv}}}$ and $s_{s_{\ell\text{cv}}}$. Using only $\ell$cv-bins with more than 144 samples (1 day of data), and only those $\ell$cv-bins where all four tests have at least 144 samples, we compute $\overline{s_{\delta_{\ell\text{cv}}}}$ and $\overline{s_{s_{\ell\text{cv}}}}$ for all bins -18 dB < CNR < -12 dB, $\ell \in [0,90,180,270,V]$ and -18 m/s < v < 18 m/s, which cover the typical CNR and wind speed conditions for profiling wind lidar. $\overline{s_{\delta_{\ell\text{cv}}}} = 0.021\ m/s$ and $\overline{s_{s_{\ell\text{cv}}}} = 0.027\ m/s$. These terms represent preliminary estimate of the reproducibility error, but it is noted here that this estimate does not meet the criteria in Formula 4 in (27).

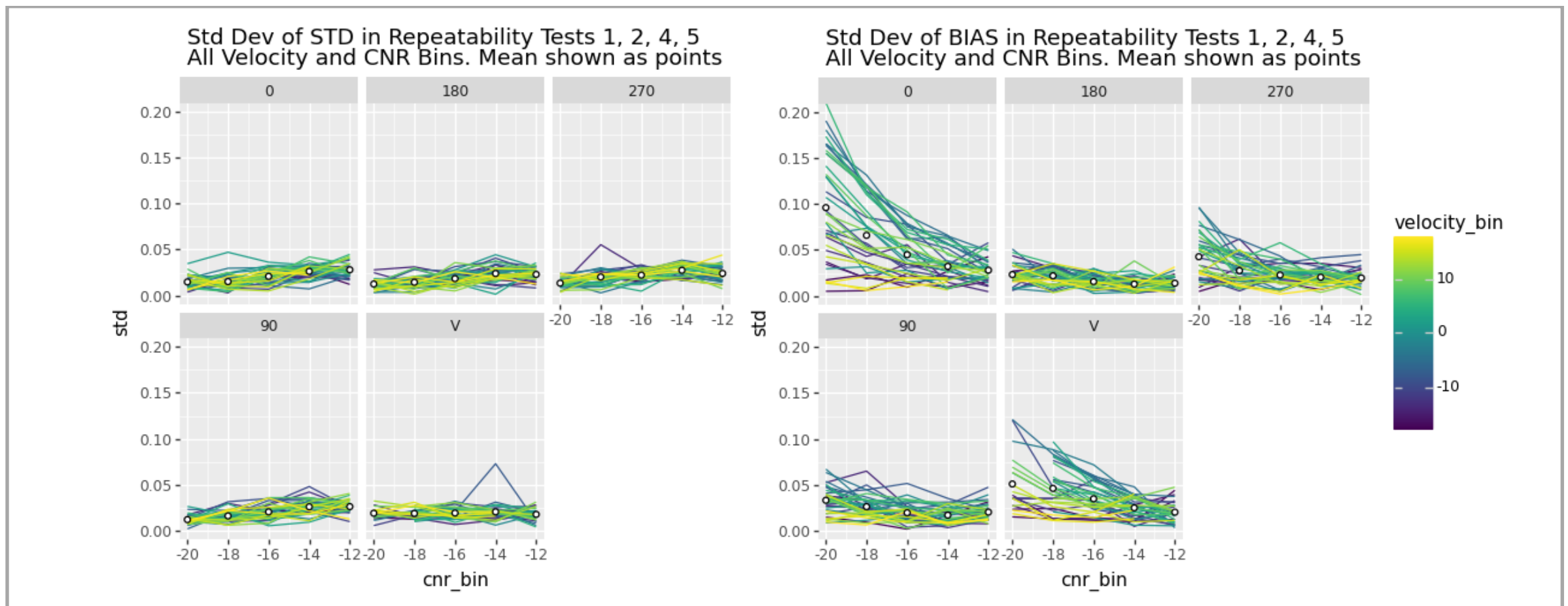

**Figure 4:** Standard deviation of measurements of standard deviation (left) and mean (right) uncertainties in each $\ell$cv-bin for the four SAFO-MV reproducibility tests.

These uncertainties are propagated though the lidar's wind field reconstruction (WFR) algorithm to derive the uncertainty of the horizontal wind speed.

Due to the subtraction of parallel beams in the WFR algorithm, fully correlated errors would exactly cancel. This is a robust aspect of the symmetrical design of the lidar. Fully anti-correlated errors are not realistic and would yield an overestimate. Therefore, it is assumed that the errors are fully uncorrelated. Propagating errors through the WFR arrives at the following expressions:

$$\delta U_{s_{\ell \mathrm{cv}}} = \frac{\overline{s_{s_{\ell \mathrm{cv}}}}}{\sqrt{2}\ sin\,\varphi} = 0.041\ m/s \tag{6}$$

For the uncertainty of the estimate of the random error, repeated sampling reduces the uncertainty by $1/\sqrt{N}$, where $N$ is the number of repeated runs.

$$\delta U_{\delta_{\ell \mathrm{cv}}} = \frac{\overline{s_{\delta_{\ell \mathrm{cv}}}}}{\sqrt{2}\sqrt{N}\ sin\,\varphi} = 0.0026\ m/s \tag{7}$$

The standard uncertainty of the SAFO-MV bench itself may be estimated in two different ways. The first is by estimating the associated uncertainty of the opto-electronic components in the bench, particularly the RF drivers (see Page 3). This method, however, is based on uncertainties of the drivers operating continuously, while the lidar under test operates in pulsed mode. A suitable alternative approach is to evaluate the uncertainty using an independent, pulsed lidar with significantly lower uncertainty than an ordinary WindCube. Following this approach, using a lidar based on the WindCube Scan 400S architecture, the bench uncertainty (k=2) for radial velocity is derived as $\delta v_{\mathrm{SAFO}} = 0.0145\ m/s$. For a typical 10-minute measurement period, each LOS samples N=150 radial velocities, yielding:

$$\delta U_{\delta v_{\mathrm{SAFO}}} = \frac{\delta v_{\mathrm{SAFO}}}{\sqrt{2}\sqrt{N}\ sin\,\varphi} = 0.0018\ m/s \tag{8}$$

MoCaLUM is able to simulate random fluctuations of the lidar beam angle. The software was first developed for scanning lidar, which use a mechanical scan head to steer the lidar beam. The scan head may exhibit random error in the lidar pointing angle. Profiling lidar use fixed telescopes, and the outbound beam trajectories do not vary randomly, but may have small, fixed errors due to manufacturing tolerances. The tolerance of the WindCube telescope orientation is 0.15°. Repeated measurements of the beam angles yield a standard deviation of 0.055°. Assuming a uniform distribution, the angle standard uncertainty based on the tolerance is reduced by $1/\sqrt{3}$ to 0.087°.

Propagating to horizontal wind speed, assuming full correlation between the errors on the four oblique telescopes yields:

$$\delta U_{\delta\varphi} = U \cot\varphi \, \delta\varphi = 0.00248 * U \; m/s \tag{9}$$

Finally, the $z_d$ calibration uncertainty affects the height at which the radial velocity is sampled, and thus can generate an error in non-uniform wind conditions. Repeated measurements of the $z_d$ timing yield a standard deviation of 3 ns, or equivalently 0.9 m. For roundtrip, backscattered light, this is reduced by half to 0.45 m. Using the typical wind shear power law:

$$U = U_0 \left(\frac{z}{z_0}\right)^{\alpha} \tag{10}$$

the error in wind speed may be derived via the error in altitude, shown here for z = 100 m and α = 0.2:

$$\delta U_{\delta z} = \alpha \, \frac{U}{z} \delta z = 0.0009 * U \; m/s \tag{11}$$

**Table 4:** Uncertainties associated with SAFO-MV bench, zero-distance calibration, and lidar geometry not modeled via Monte Carlo approach. Values shown for z = 100 m and α = 0.2. Total row is sum, in quadrature, of all other terms

| Term | $U$=4 m/s | 6 | 8 | 10 | 12 | 14 | 16 |
|---|---|---|---|---|---|---|---|
| $\delta U_{s_{\ell cv}}$ | 0.041 | 0.041 | 0.041 | 0.041 | 0.041 | 0.041 | 0.041 |
| $\delta U_{\delta_{\ell cv}}$ | 0.0026 | 0.0026 | 0.0026 | 0.0026 | 0.0026 | 0.0026 | 0.0026 |
| $\delta U_{\delta v_{SAFO}}$ | 0.0018 | 0.0018 | 0.0018 | 0.0018 | 0.0018 | 0.0018 | 0.0018 |
| $\delta U_{\delta\varphi}$ | 0.0099 | 0.0149 | 0.0198 | 0.0248 | 0.0298 | 0.0347 | 0.0397 |
| $\delta U_{\delta z}$ | 0.0036 | 0.0054 | 0.0072 | 0.009 | 0.0108 | 0.0126 | 0.0144 |
| ***Total (m/s)*** | 0.0424 | 0.0441 | 0.0462 | 0.0489 | 0.0519 | 0.0553 | 0.0589 |
| ***Total (%)*** | 1.06 % | 0.74 % | 0.58 % | 0.49 % | 0.43 % | 0.40 % | 0.37 % |

The uncertainties in Table 4 are added in quadrature to the uncertainties derived from the Monte Carlo simulation. These represent the inherited uncertainty of the reference instrument, as well as other uncertainties which are not properly modeled by the Monte Carlo simulation

## 3. Results

Two WindCube lidars were evaluated following the methodology described above and compared to calibrations and classifications following the IEC 61400-50-2. In this comparison, the RWFs from systems WLS7-1100 (WindCube v2.1) and WLS7-9982 (WindCube XP) were retrieved from the manufacturing traceability database and used in the MoCaLUM simulation. For system WLS7-1100, which was manufactured and classified in 2019, before development of the SAFO-MV bench. there are no SAFO-MV calibration records. Instead, the SAFO-MV calibration of a different WindCube v2.1 system, WLS866-0195, manufactured in 2023 is used to represent that device's behaviour.

### 3.1. SAFO-MV Calibration Results

Using the SAFO-MV bench, following the sweep parameters in Table 1, WindCube v2.1 and WindCube XP systems were calibrated. WLS866-0195 is a WindCube v2.1. WLS7-9982 is a WindCube XP. These are different model versions but have identical beam angles and other common elements than allow for the SAFO-MV bench to operate identically on each system. After gathering the 1.25 Hz data (real-time data or "RTD") from the devices and the SAFO-MV control system, the data were binned by CNR, LOS, and the programmed velocity shift. The calibration for WLS866-0195 started on February 17, 2023, 16:07 UTC and ended on February 20, 2023, 1:05 UTC, lasting 57 hours. The calibration for WLS7-9982 started on December 1, 2023, 15:31 UTC and ended on December 2, 2023, 23:59 UTC, lasting 33 hours. For each bin, the mean and standard deviation of the difference between the programmed shift and the measured velocity was computed. These are shown in Figure 5. The calibration results are used to simulate the random instrumental noise in simulated radial velocity measurements in the MoCaLUM simulation. For each virtual radial measurement, a random value drawn from a Gaussian distribution is added to the radial velocity. The mean and standard deviation of the distribution are selected from the SAFO calibration results according to the mean radial wind speed, the LOS, and the configured CNR of the simulation.

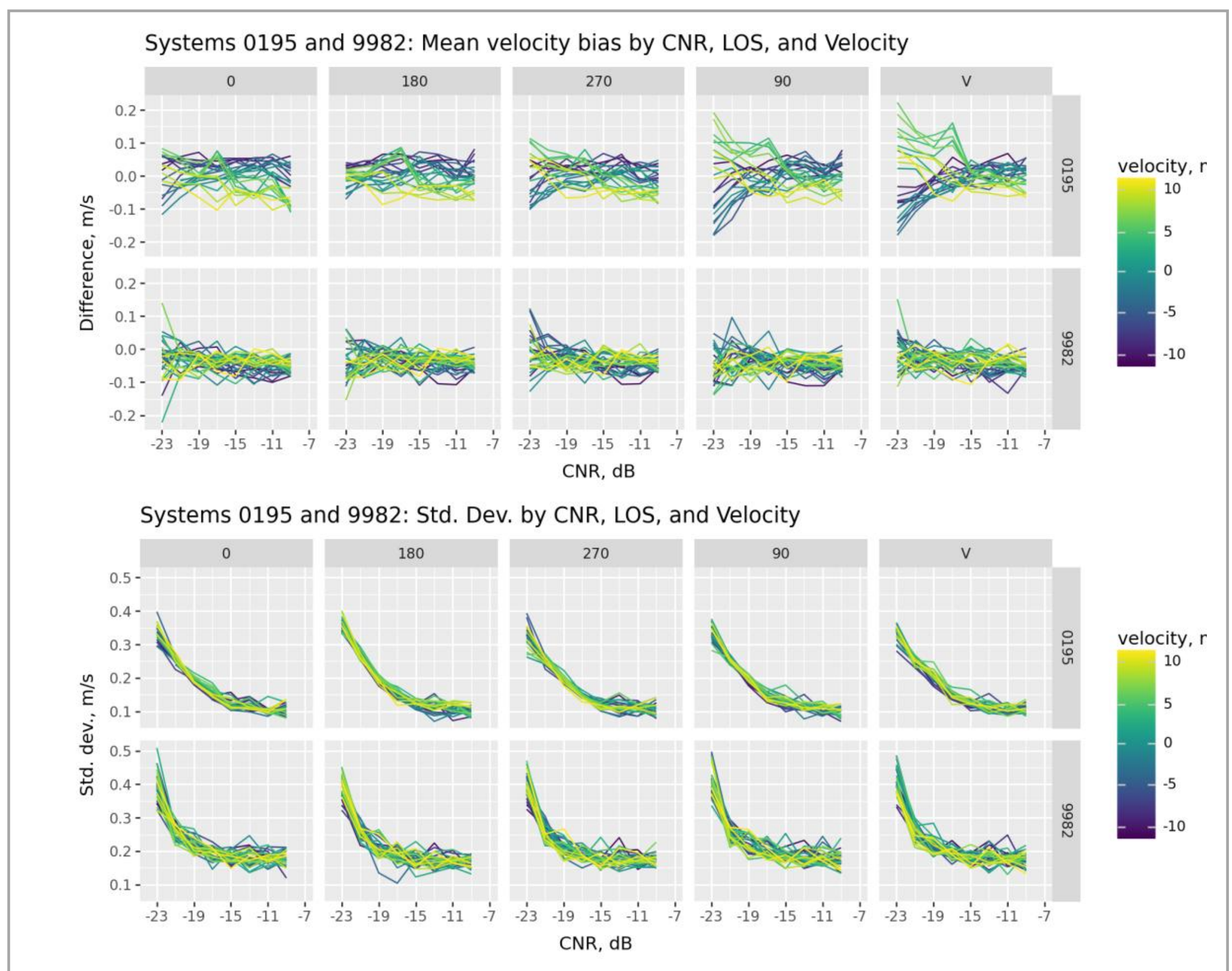

**Figure 5:** (top) Mean difference between measured radial velocity and programmed velocity shift for two WindCube variants. (Bottom) Standard deviation of measured radial velocity. WLS866-0195 is a marinized WindCube v2.1 manufactured in 2023 and WLS7-9982 is an onshore WindCube XP

### 3.2. Zero-distance Calibration and Range-Weighting Function Measurement

The pulse, apodization function, and RWF are shown in Figure 6. The apodization functions are different due to a change in the number of points in the Fast Fourier Transform (64 to 128) between versions v2.1 and XP.

### 3.3. MoCaLUM Derivation of Environmental Sensitivities

In MoCaLUM, the RWF and SAFO-MV control the virtual lidar instrument behaviour. The input wind direction, wind speed, and CNR determine which distribution is used to generate random errors. The ambient wind conditions are a time series of observed conditions during the classification tests of each system type. The details of the type classification measurement campaigns are shown in Table 5. For each 10-minute measurement period, the wind shear is computed at each altitude using the Equation 10 and the measured wind speeds at the two nearest neighbor heights, and one neighbor for the lowest and highest measurement heights. The CNR gradient is computed in the same manner. Turbulence intensity is computed from the lidar data. All inputs to the PyConTurb turbulence box are listed in Table 2.

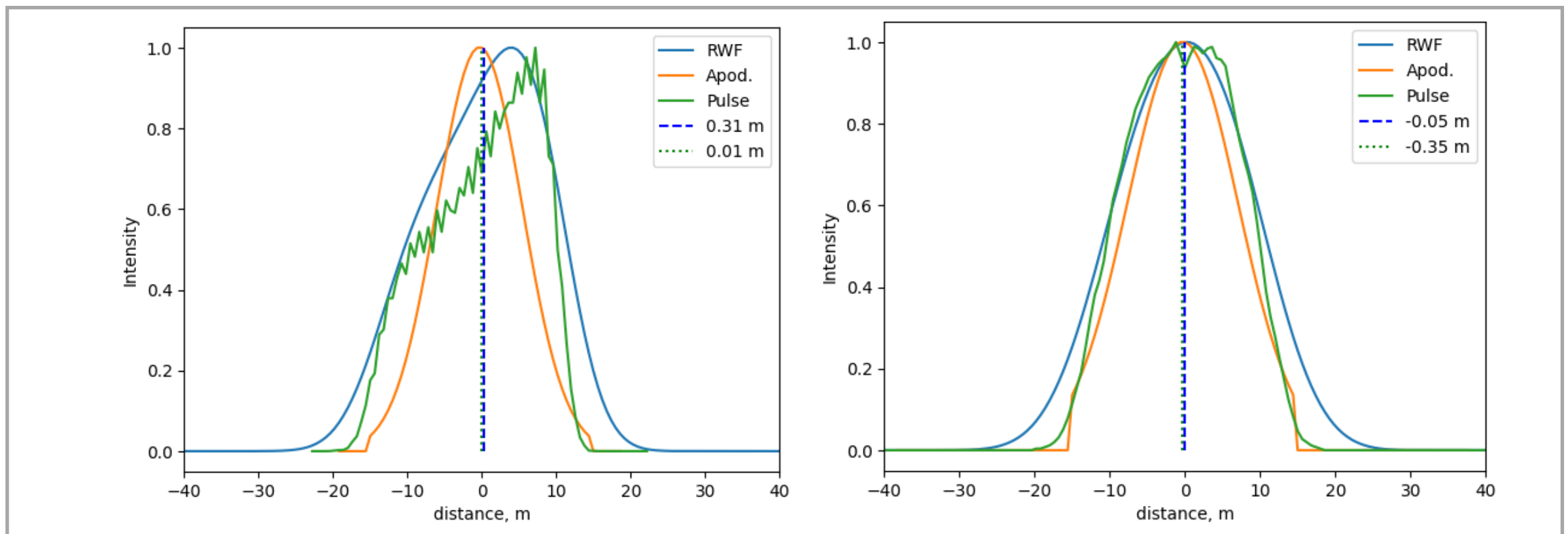

**Figure 6:** Lidar pulse, apodization function, and final range-weighting function for (left) system WLS7-1100 and (right) system WLS7-9982, with barycenter of pulse and RWF noted. The lidar $Z_d$ is computed using the barycenter of the full RWF.

In remote sensing device type classification tests following the IEC 61400-50-2, the device sensitivities to environmental variables are derived by:

- Binning the data by the specified environmental variable (e.g. wind shear)
- Computing the mean difference between the lidar and the reference anemometer
- Filtering out bins that do not contain sufficient sample number
- With the remaining bins, compute a two-factor linear regression with the environmental variable as the independent variable, and the differences as the dependent variable

The regression must exceed the threshold of one of two significance tests:

$$|m * std| > 0.5\ \% \qquad (12)$$

$$|m * std * R| > 0.1\ \% \qquad (13)$$

Where $m$ is the derived slope, $std$ is the standard deviation of the environmental variable, and $R$ is the correlation coefficient of the regression. The classification of WindCube v2.1 identified wind shear, turbulence intensity, and wind direction as sensitivities. The WindCube XP classification identified wind shear as a sensitive variable. Using MoCaLUM, the wind shear and turbulence intensity sensitivities are examined for the specific systems WLS7-1100 and WLS7-9982 (28) (29) (30).

**Table 5:** WindCube type classification measurement campaign detail. MoCaLUM input conditions for WLS7-1100 are taken from the Hamburg campaign, and for WLS7-9982 from the Rodewald campaign.

| Device Type | Serial Number | Start Date | End Date | Site Name | Altitudes |
|---|---|---|---|---|---|
| WindCube v2.1 | WLS7-1100 | 2020/2/20 | 2020/6/10 | Hamburg | 45, 65, 80, 121 |
| WindCube v2.1 | WLS7-1100 | 2020/2/20 | 2020/6/10 | Georgsfeld | 40, 60, 82, 100, 120, 131, 135 |
| WindCube v2.1 | WLS7-1260 | 2020/4/24 | 2020/12/3 | Georgsfeld | |
| WindCube XP | WLS7-9982 | 2024/8/11 | 2024/11/17 | Rodewald | 80, 100, 120, 140, 160, 180, 200 |
| WindCube XP | WLS7-9982 | 2024/2/22 | 2024/5/31 | Janneby | 60, 80, 100, 120, 140, 160, 180, 200 |
| WindCube XP | WLS7-9983 | 2024/2/22 | 2024/5/31 | Janneby | |

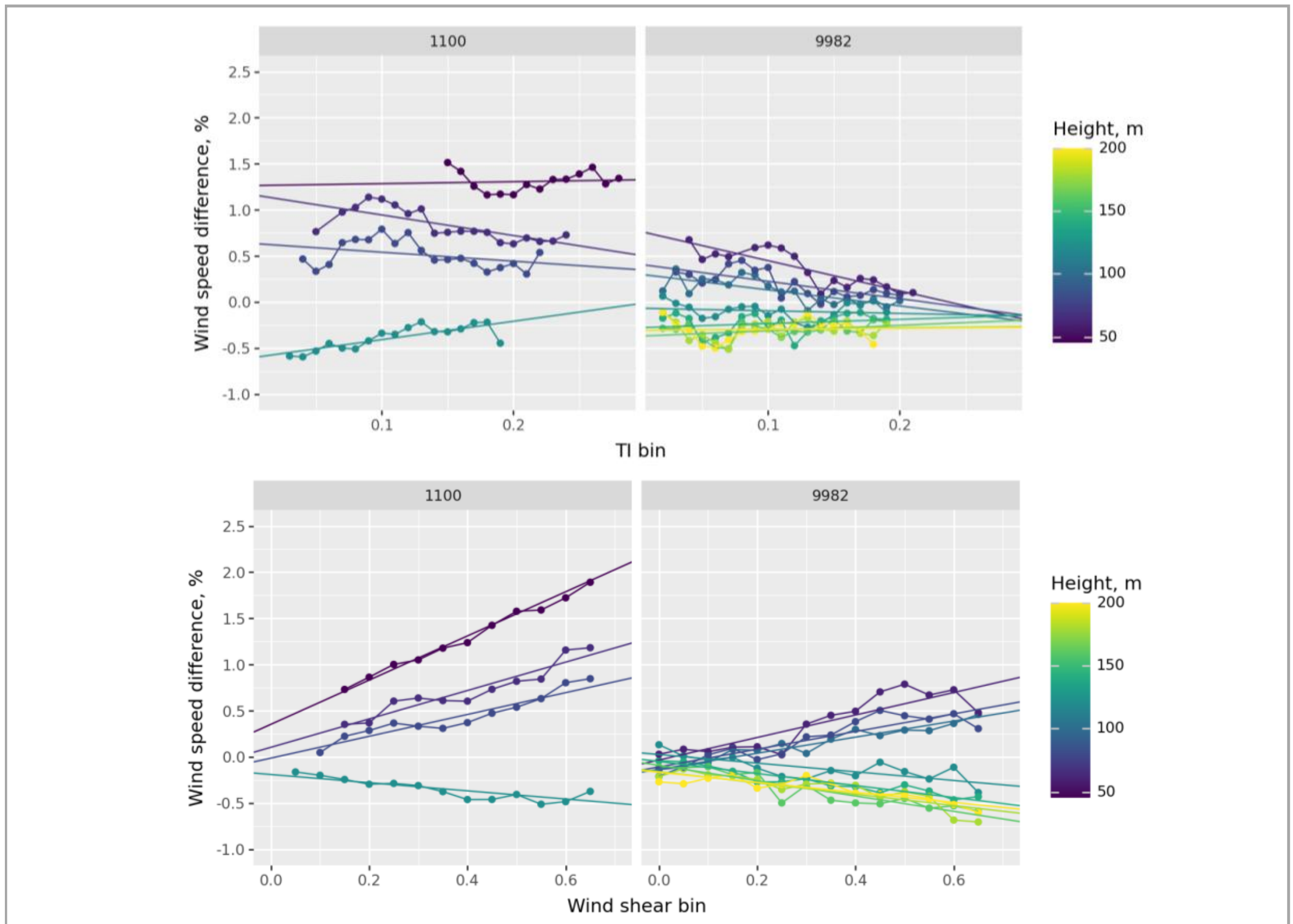

**Figure 7:** MoCaLUM-simulated binwise differences between virtual lidars (WLS7-1100 and WLS7-9982) and virtual reference for turbulence intensity (top) and wind shear (bottom). Input conditions are set by the time series of measurements the type classifications described in Table 5 (Hamburg for WLS7-1100 and from Rodewald for WLS7-9982) and other input conditions are described in Table 2. Regression slopes are shown as solid lines.

In Figure 8 the derived sensitivity slopes are compared to the sensitivity slopes calculated from the classification measurement campaigns for the specific systems at both sites.

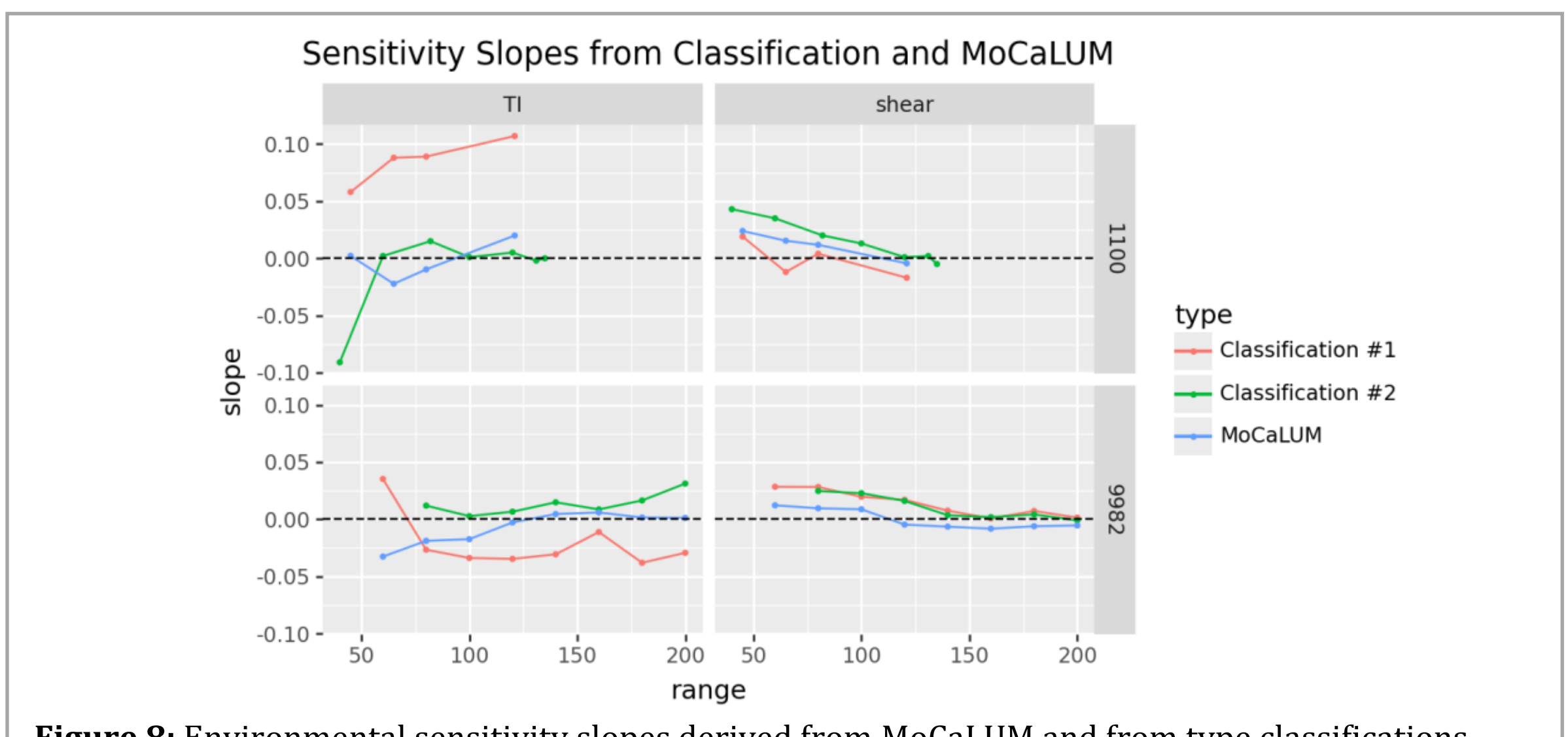

**Figure 8:** Environmental sensitivity slopes derived from MoCaLUM and from type classifications. Classification #1 for WLS7-1100 is from Hamburg, #2 from Georgsfeld. Classification #1 for WLS7-9982 is from Rodewald and #2 from Janneby.

The sensitivity slopes for wind shear agree quite well for system WLS7-1100. The simulated slopes fall in between those observed at the two classification sites, and the simulated slopes exhibit a negative trend with altitude of similar magnitude to the Classification slopes. For system WLS7-9982, the simulated wind shear sensitivity slopes exhibit the same negative trend with altitude, and the same total change in slope between 60 m and 200 m. However, the simulated slopes show a zero-crossing at 110 m, while the Classification slopes approach zero around 140 m, but do not cross to consistent negative values.

The simulated turbulence sensitivity slopes for WLS7-1100 agree well with the derived slopes from the Janneby site, except at 45 m, where the observed slope has a large negative value. WLS7-1100's simulated slopes disagree with those observed at the Hamburg site, which are almost 10 times larger than the simulated slopes, and those observed at Janneby. For WLS7-9982, the observed turbulence sensitivity slopes fall between those observed at Rodewald and Janneby.

### 3.4. MoCaLUM Calibration Uncertainty

Lidar calibration following the IEC 61400-50-2 computes wind speed uncertainty in 0.5 m/s wind speed bins. In each wind speed bin, the uncertainty components are:

- The standard uncertainty of the reference sensor
- The mean difference between the lidar and the reference sensor
- The standard uncertainty of the lidar measurement, $\sigma_{WS}/\sqrt{N}$, for $N$ samples in the bin
- The uncertainty of the lidar due to its mounting (often 0 %)
- The uncertainty due to the distance of the lidar and the refence mast
- The uncertainty associated with inhomogeneous flow through the lidar measurement volume (often 0 %)

In the IEC calibration, the largest uncertainty is the standard uncertainty of the reference sensor, a calibrated cup anemometer mounted on a reference mast. This component is shown in Figure 9 as "Calibration" for each approach. In the MoCaLUM time series calibration approach, where measured data are used as inputs to the simulation, the uncertainty terms in each wind speed bin are:

- The standard uncertainty of the virtual reference sensor (see Table 4)
- The mean difference between the virtual lidar and the virtual reference sensor
- The standard deviation of the difference between the virtual lidar and virtual reference
- The standard uncertainty of the difference $\sigma_{diff}/\sqrt{N}$, for $N$ samples in the bin

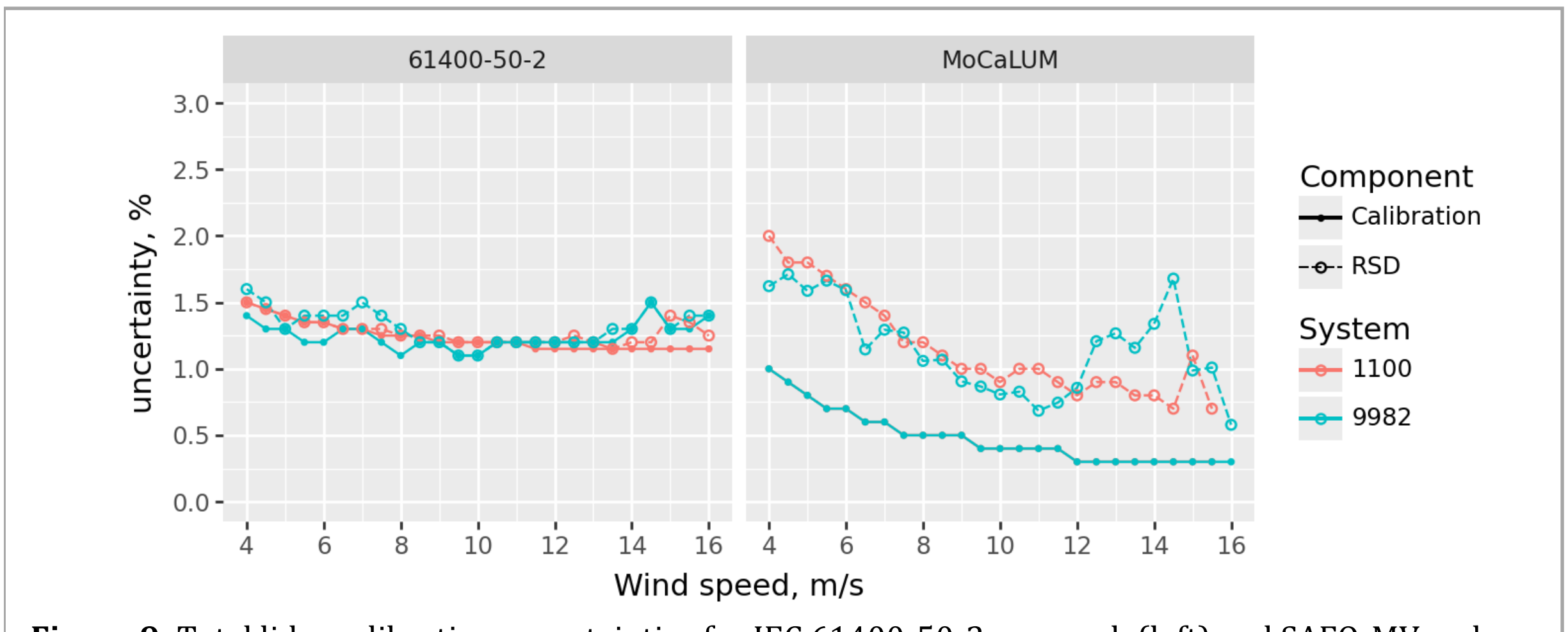

**Figure 9:** Total lidar calibration uncertainties for IEC 61400-50-2 approach (left) and SAFO-MV and MoCaLUM approach (right) for 121 m (WLS7-1100) and 120 m (WLS7-9982) Wind speed difference and standard deviation for virtual lidar for all altitudes shown in Appendix.

A comparison of the two different approaches is shown in Figure 9. For both approaches, the two lidar systems agree closely for wind speeds between 4 m/s and 12 m/s. Above 12 m/s, low sample number for WLS7-9982 wind speed bins leads to an increase in the calibration uncertainty for both methods. For the IEC approach, this increase is barely visible above the inherited reference uncertainty, while for the MoCaLUM approach, this increase is more pronounced. IEC calibration uncertainties are often identical to the inherited uncertainty from the reference instrument, shown as filled circles in Figure 9. Conversely, the MoCaLUM-derived uncertainties are always significantly higher than the reference uncertainty. For the IEC method, the uncertainties are between 1.1 % and 1.6 %. For the MoCaLUM method, the uncertainties range between 0.5 % and 2.0 %. For low wind speeds, MoCaLUM yields higher uncertainties than the IEC method, while above 8 m/s MoCaLUM uncertainties are lower than the IEC uncertainties, except in cases of low sample number noted above.

## 4. Discussion

The presented results offer a novel procedure to calibrate ground mounted lidars and derive measurement uncertainties of horizontal wind speed during a specific measurement campaign (SMC). For wind energy applications, the IEC 61400-50 series standards describe wind measurement uncertainties based on applied measurement technology. In this discussion, the standards for cup and sonic anemometry (10), ground mounted lidar (1) and nacelle mounted lidars (2) are of interest . Common to these standards is the need for a calibration of each individual instrument unit and a type-classification assessing operational characteristics. However, the specific methods on performing these tasks differ between the standards.

For ground mounted lidar, the uncertainties, described in the 50-2 standard are derived via field calibration of horizontal wind speed against anemometers mounted on measurement masts ("met masts") and device classification to assess sensitivities to environmental variables, also using met masts as a reference. Due to this calibration method, the lidar inherits the uncertainties associated with the anemometers. These include those originating from calibration in wind tunnels, operational characteristics of the anemometer, mounting on the met mast, and the flow characteristics of the calibration site. In addition, the height of the met mast limits the measurement range which can be calibrated and classified.

The field-test classification method of Chapter 6 of 50-2 is quite limited. Observed deviations in horizontal wind speeds measured by anemometers, and wind speeds measured by lidar are analysed as function of any environmental data expected to have an impact on the lidars measurement output. This only allows for sampling the environmental conditions available at the test site and only with correlations between the variables as present in the measurement period. This undersampling of the parameter space is addressed by favouring in the calculation of the measurement uncertainty based on sensitivity slopes and the actual difference in conditions between calibration and SMC.

In contrast, cup anemometers are calibrated in wind tunnels with full control of the horizontal wind speed (10). The classification on the other hand is derived with the application of a numerical anemometer model. The model's input parameter concerning the anemometer properties like rotor and friction torque, response to inflow inclination, to name but a few, are experimentally determined in wind tunnel tests (Chapter 7.2 of 50-1). Given these instrument properties, the anemometer's response to an artificial wind flow is calculated for the full parameter space of environmental ranges as defined in Table 1 of 50-1 (about $10^5$ simulations in practice). The maximum systematic deviation from a wind tunnel calibration defines the class number, which then is used as foundation of the uncertainty calculation.

Common to the approaches of 50-1 and 50-2 is that the calibration is made with the parameter of interest, i.e. the horizontal wind speed. For nacelle-mounted lidars, the horizontal measurement geometry makes it very challenging to reliably calibrate the horizontal wind speed. The IEC 61400-50-3:2022 standard therefore defines a different approach: the lidar's derivation of horizontal wind speed is split into first, the measured, intermediate values, and second, the WFR. Intermediate values are line-of-sight wind speed, beam geometry and instrument inclination (tilt and roll). Only the intermediate values are calibrated. For the LOS wind speed, a calibration against a met mast and a sensitivity analysis based on 50-2 are performed and the resulting measurement uncertainties propagated through the WFR. This propagation can be done either according to JCGM 100:2008 (GUM) or JCGM 101:2008 (Monte Carlo). In addition, the adequacy of the WFR for the environmental condition of the SMC shall be tested in field tests by showing that the WFR does result in comparable results to a met mast-based measurement campaign. This last step limits the possible applications to flat terrain and offshore.

Both the 50-2 and 50-3 lidar standards do not include alternative methodologies without the usage of cup anemometers for deriving traceable uncertainties, as such alternative methods had not reached maturity at the time the standards were written. The use of SAFO-MV and MoCaLUM to derive traceable lidar uncertainties is one of several novel methods in development in the wind energy industry and academy. Others include those presented in (8) and (9) which both use specialized lidar as reference instruments.

In the context of the above summarised calibration and classification schemes, the SAFO-MV and MoCaLUM procedure utilises aspects of multiple standards: the control of ambient conditions and duration of SAFO-MV bench calibration is comparable to that of an anemometer calibration in a wind tunnel. By experimentally testing intermediate values via SAFO-MV instead of the horizontal wind speed, the new approach follows concepts found in the 50-3. Error propagation by Monte Carlo simulation is directly mentioned in 50-3. The key differences are the types of uncertainty components explicitly mentioned in 50-1, 50-2 and 50-3, which arise from the specific reference instruments, either wind tunnels or cup anemometers. SAFO-MV is new, and thus includes several components not found in the literature.

The MoCaLUM procedure yields uncertainties very similar to those derived following the IEC 61400-50-2. The observed increases in uncertainty at low speeds originate from the direct inclusion of the standard deviation, rather than only the standard uncertainty. The increased range of the final calibration values from MoCaLUM (0.6 % - 2.0 %) compared to the IEC method (1.1 % - 1.5 %) is due to the reduction in inherited uncertainty from the reference instruments. As such, this methodology is more sensitive. The IEC calibration method takes a long time. The datasets used in this study took a minimum of 95 days, though these campaigns were designed for classification rather than calibration. Calibration campaigns typically take about 45 days to fill the necessary wind speed bins, and this duration depends on the seasonality of the wind at the calibration site. SAFO-MV calibration takes as few as 33 hours to complete. The MoCaLUM simulation takes an additional 1-6 hours depending on the processor speed of the computer. Including data collection, verifying proper configuration, and data quality control, the entire calibration process takes roughly 48 to 72 hours, more than a 90 % reduction in the calibration time, while delivering a similar uncertainty range, with reduced reference uncertainty.

The observed environmental sensitivities from the IEC classification tests are well replicated in the simulation environment, except for the turbulence sensitivity of WLS7-1100 at the Hamburg test site. It is possible, but not proven, that the field calibration sensitivity is an artifact of the anemometer or the site. These types of anomalous results are not uncommon in lidar field classifications and calibrations, and are a key reason why laboratory and simulated environments are preferable for lidar calibration. The general agreement between the simulated and field-derived environmental sensitivities, as well as the good alignment of calibration uncertainties constitute a validation of the proposed measurement model.

The usage of the MoCaLUM is not limited to the propagation of calibration uncertainties through the WFR. It also enables derivation of uncertainties covered by type classification. Here MoCaLUM offers several opportunities when assessing the classification uncertainty:

a) compile a maximum uncertainty for a given set of environmental conditions by sampling the full parameter space like it is done in 50-1,
b) derive sensitivity slopes for the variables like the 50-2
c) perform the uncertainty propagation for a specific measurement campaign.

Like the anemometer model examples given in 50-1, the MoCaLUM approach cannot capture components that are not explicitly put into the model. One parameter is the impact of flow inhomogeneity in complex terrain. As these impacts are not assessable with the current classification scheme according to 50-2, the IEC 61400-12-1 excludes complex terrain. However, it is well accepted today that this is assessed adequately with other methods, taking CFD models as an example (31). As those assessments are already performed independently from calibration and classification results, the introduction of the SAFO-MV and MoCaLUM method does not impact these.

## 5. Conclusions

In this research, a new methodology for calibration of Doppler wind lidar is validated via comparison with existing IEC methods, and successful demonstration for two different product generations of the device class. The new method uses a Monte Carlo approach, relying on calibration of the lidar LOS using an opto-electronic bench, and a virtual lidar embedded in a KSEC turbulence box for derivation of the final uncertainties. Using the same toolchain, the devices' environmental sensitivities are derived and shown to be in general agreement with those observed in IEC field classifications. The calibration uncertainties derived using the Monte Carlo method have a slightly larger range than those from the IEC method. This increased range arises from differences in the included uncertainty terms, as well as the reduction of inherited uncertainty from the reference instruments. These results, demonstrated for two different versions of the same lidar, constitute a validation of the measurement model, and are a key step for bringing the new method into industrial use via inclusion to IEC standards. The new approach reduces the total time to complete an SI-traceable calibration for Doppler lidar from roughly 1.5 months to less than three days, a 90 % reduction in calibration time. The method demonstrated in this study, namely the use of a measured time series as the simulation input conditions, is not the only possible embodiment of this approach for industrial lidar calibrations. Further study and input from industry stakeholders and expert groups will inform the exact implementation.

## Acknowledgements

The authors would like to acknowledge Ørsted, whose support of SAFO-MV has been instrumental to ongoing efforts to industrialize the bench. Andrea Vignaroli, Jochen Rainier Cleve, and Jana Preissler contributed significant new modules to Nikola Vasiljević's original MoCaLUM codebase. Azzurra Bigioli and Clarise Vellosso developed the SAFO-MV uncertainty budget and painstakingly validated each component. Florestan Ogheard gave Vaisala critical guidance on uncertainty and SI-traceability.

# 6. Appendix

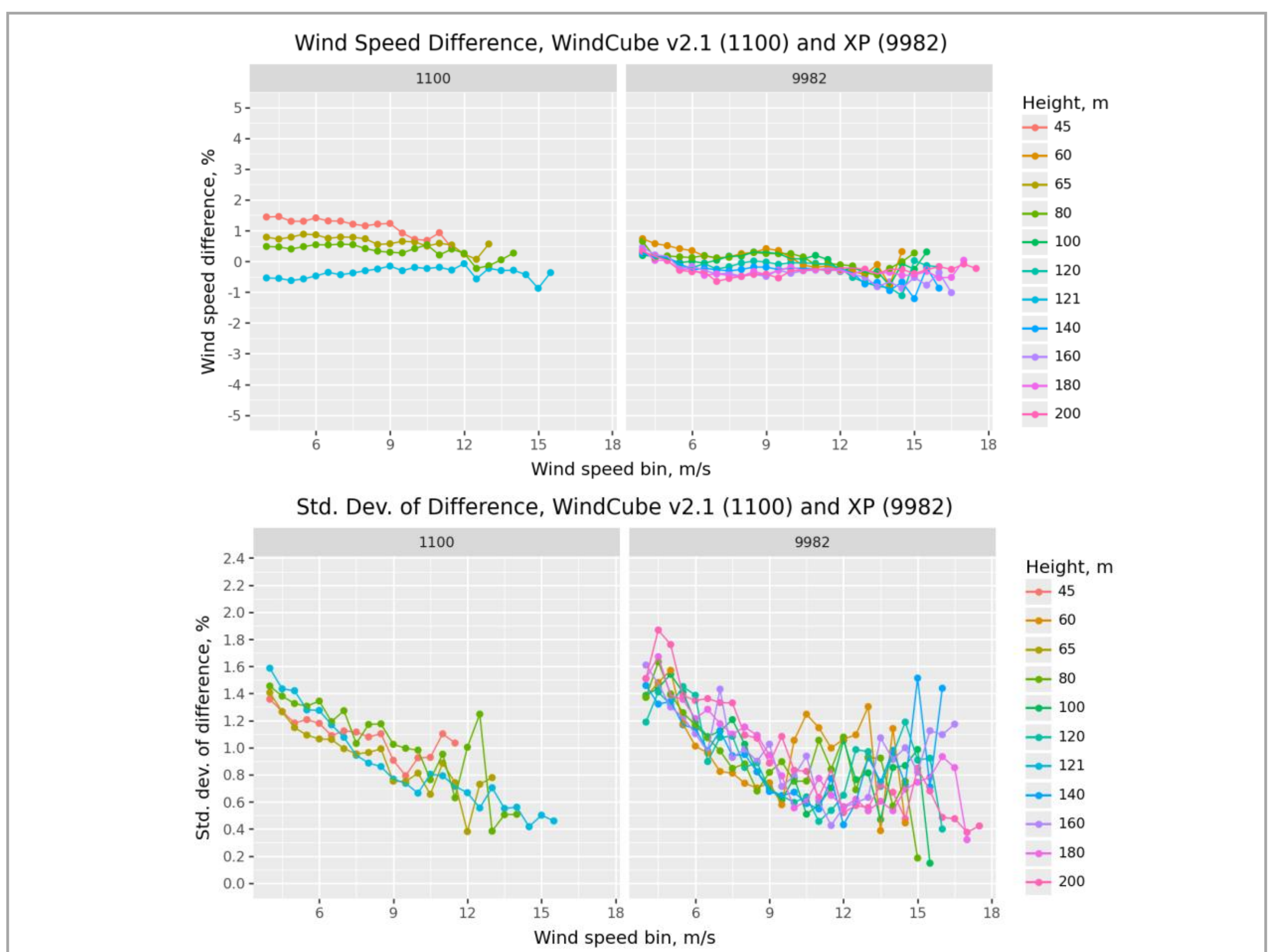

**Figure 10:** (Top) Wind speed difference between virtual lidar and virtual refence at different altitudes for each system by wind speed bin (Bottom) Standard deviation of wind speed difference between virtual lidar and virtual reference for each wind speed bin.